\documentclass[12pt]{article}

\usepackage{times}
\usepackage{color}
\usepackage[bottom]{footmisc}
\usepackage{hyperref}
\usepackage{graphicx}
\usepackage{epstopdf}
\usepackage{dcolumn}
\usepackage{bm}
\usepackage{mathtools}

\newcounter{lastnote}

\title{Longitudinal patterning of twisted light}

\author
{Ahmed H. Dorrah$^{1\ast}$, Michel Zamboni-Rached$^{2}$, and Mo Mojahedi$^{1}$\\
\\
\normalsize{$^{1}$Edward S. Rogers Sr. Department of Electrical and Computer Engineering,}\\
\normalsize{ University of Toronto, Toronto, ON, M5S 3G4, Canada.}\\
\normalsize{$^{2}$State University of Campinas, Campinas, SP, Brazil.}\\
\\
\normalsize{$^\ast$E-mail: Ahmed.Dorrah@mail.utoronto.ca}
}

\date{}

\begin{document} 


\baselineskip24pt


\maketitle


\textbf{Light beams with azimuthal phase dependence [$exp(i \ell\phi)$] carry orbital angular momentum (OAM) which differs fundamentally from spin angular momentum (SAM) associated with polarization. Striking difference between the two momenta is manifested in the allowable values: where SAM is limited to $\hbar k_0$ per photon, the OAM has unbounded value of $\ell\hbar$ per photon ($\ell$ is integer), thus dramatically exceeding the value of SAM \cite{Ref1,Ref2, Ref3}. OAM has thus been utilized in optical trapping \cite{Ref4}, imaging\cite{Ref2}, and material processing \cite{Ref5}. Furthermore, the unbounded degrees-of-freedom in OAM states have been deployed in data communications \cite{Ref6}. Here, we report an \textit{exceptional} behavior for a class of light beams---known as Frozen Waves (FWs)---whose intensity and azimuthal phase profiles can be controlled along the propagation direction, at will. Accordingly, we generate rotating light patterns that can change their sense of rotation and order of phase twist with propagation. Manipulating OAM along the beam axis can open new directions in optical science and its applications.}  

Twisted light beams (optical vortices) are characterized by azimuthal phase dependency $exp(i\ell\phi)$. As such, they posses helical phase-fronts along the beam axis. The sign and value of $\ell$---referred to as the topological charge---define the handedness and order of the light twist (helicity), respectively \cite{Ref6a}. OAM can be classified as intrinsic or extrinsic \cite{Ref6b}. Upon light interaction with small particles, extrinsic OAM induces particle rotation about the beam axis, whereas intrinsic OAM causes particle rotation about its center. Examples of OAM beams are the Laguerre-Gaussian and Bessel-Gauss beams \cite{Ref2,Ref3}.   

Interestingly, when vortex beams (OAM states) of opposite topological charges are superimposed, light structures that rotate along the beam axis can be generated \cite{Ref7,Ref8,Ref9,Ref10,Ref11,Ref12}. The rotating structures can carry non-zero OAM \cite{Ref13} or zero \textit{global} OAM \cite{Ref14,Ref15,Ref16}; whereas in some other cases the rotating structure can exhibit accelerated rotation \cite{Ref17}. In all previous cases, the OAM beams carried a fixed topological charge ($\ell$) and hence the phase-fronts exhibited the same handedness and order of twist along the beam axis. The ability to control the topological charge (i.e. order of twist) of OAM states along the beam axis \textit{has not been realized}.

Here, we demonstrate advanced control of OAM by using a class of beams known as FWs. These waveforms possess localized intensity profiles that can be predesigned along the beam axis \cite{Ref18,Ref19,Ref19c}, qualifying them for applications in optical trapping and micro-manipulation \cite{Ref19a,Ref19b}. Furthermore, we demonstrate a novel feature of FWs, namely controlling the sign and value of the topological charge ($\ell$) along the beam's propagation direction. With this level of control, the helical phase-fronts of the beam can be manipulated along the beam axis, thus reversing its handedness with propagation. Moreover, by superposing several FWs (OAM states) with different topological charges, we generate rotating intensity patterns that change their sense of rotation and topological order with propagation, on-demand. These additional degrees-of-freedom can be utilized in data communications, spectroscopy, optical trapping and micro-manipulation, to name a few.

Frozen Waves [$\psi(\rho,\phi,z,t)$] can be engineered in arbitrary linear, isotropic, and homogeneous medium by performing superposition of equal frequency co-propagating Bessel beams with the same order ($\ell$). The theoretical foundation of FWs has been presented in Ref. \cite{Ref18,Ref19,Ref19c} and further developed in \cite{Ref20}. In addition, FWs with several longitudinal intensity profiles have been experimentally generated in \cite{Ref21,Ref22,Ref22b}. Here, we show advanced OAM control in waveforms generated by adding multiple FW states ($\psi_{\ell}$) such that
\begin{equation}
\label{Eq1}
\Psi(\rho,\phi,z,t) = \sum_{\ell=-\infty}^{\infty}\psi_{\ell} = e^{-i\omega t} \sum_{\ell=-\infty}^{\infty}\sum_{m=-N}^{N} A_{\ell m} J_{\ell}(\eta_{\ell m} \rho) e^{i\ell \phi} e^{i\zeta_{\ell m}z}.
\end{equation}
Each FW state $\psi_{\ell}$ is composed of $2N+1$ Bessel beams of equal order ($\ell$), thus representing the same OAM mode. For the $m^{th}$ Bessel beam in each FW state, the transverse wavenumber $\eta_{\ell m}$ is related to the longitudinal wavenumber $\zeta_{\ell m}$ via the consistency relation  $\eta_{\ell m} = \sqrt{k^2 - \zeta_{\ell m}^2}$ (conditions on $\eta_{\ell m}$ and $\zeta_{\ell m}$ can be found in the Methods section). The complex coefficients $A_{\ell m}$ represent weighting factors for the Bessel beams in the superposition such that
\begin{equation}
\label{Eq4}
A_{\ell m} = \frac{1}{L}\int_{0}^{L} F_{\ell}(z) e^{-(i\frac{2\pi}{L}m)z}dz,
\end{equation} 
where the morphological function $F_{\ell}(z)$ defines the desired longitudinal intensity profile of each FW state ($\psi_{\ell}$) over a finite distance $L$. Performing a superposition of multiple FW states that carry different topological charges ($\ell$), via Eq.~\ref{Eq1}, is a novel feature \textit{introduced here for the first time}. By properly adjusting $F_{\ell}(z)$, the contributions of each FW state $\psi_{\ell}$ (OAM mode), along the beam axis, can be determined at will. As such, specific FW state(s) with desired topological charge ($\ell$) can be selected to effectively contribute to the beam over finite space interval while other states are readily \textit{switched off}. The \textit{effective} topological charge ($\ell$) can thus be controlled along the beam axis. 

\begin{figure}[h!]
	\centering
	\includegraphics[width=0.5\textwidth]{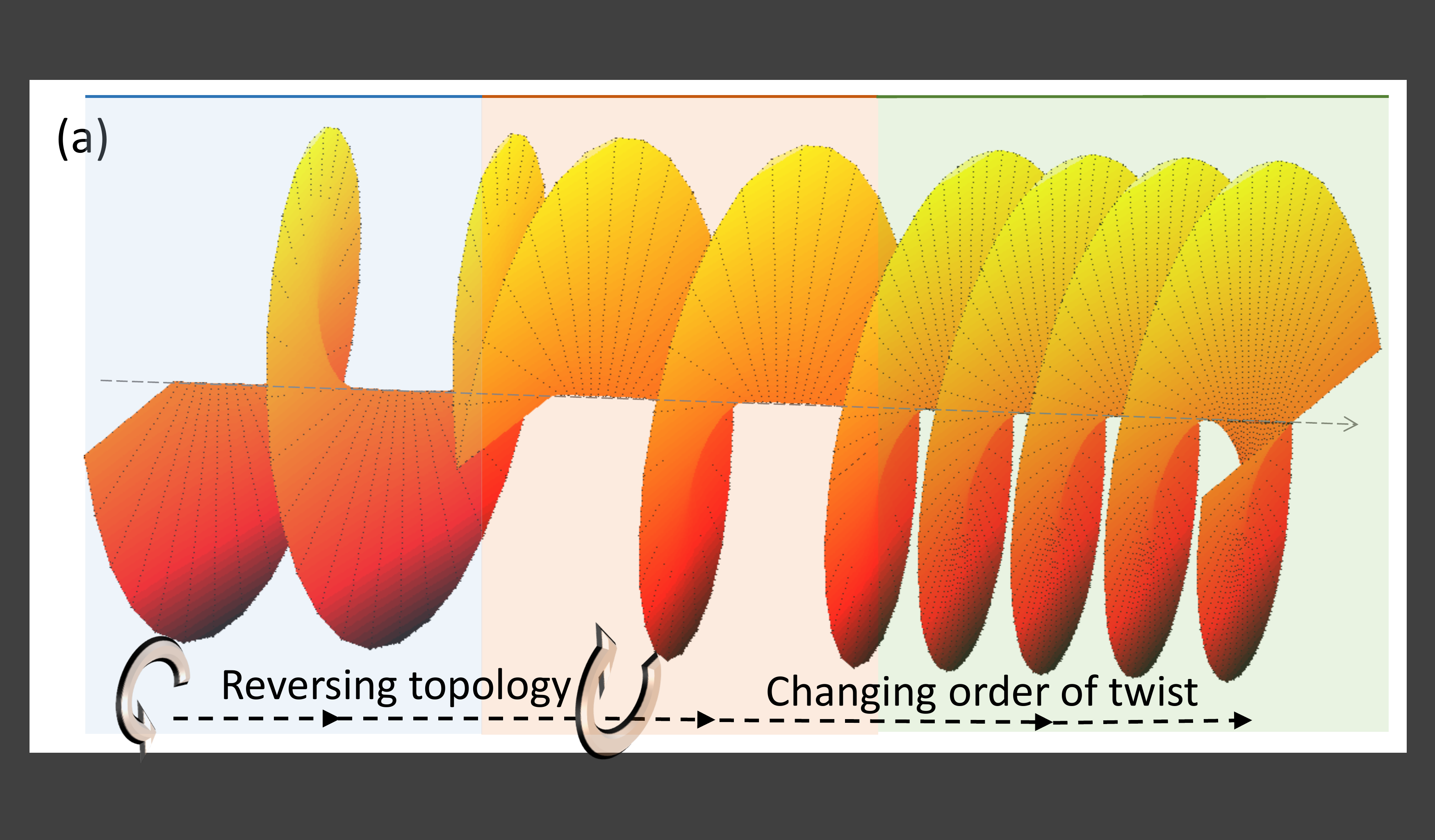}
	\includegraphics[width=0.5\textwidth]{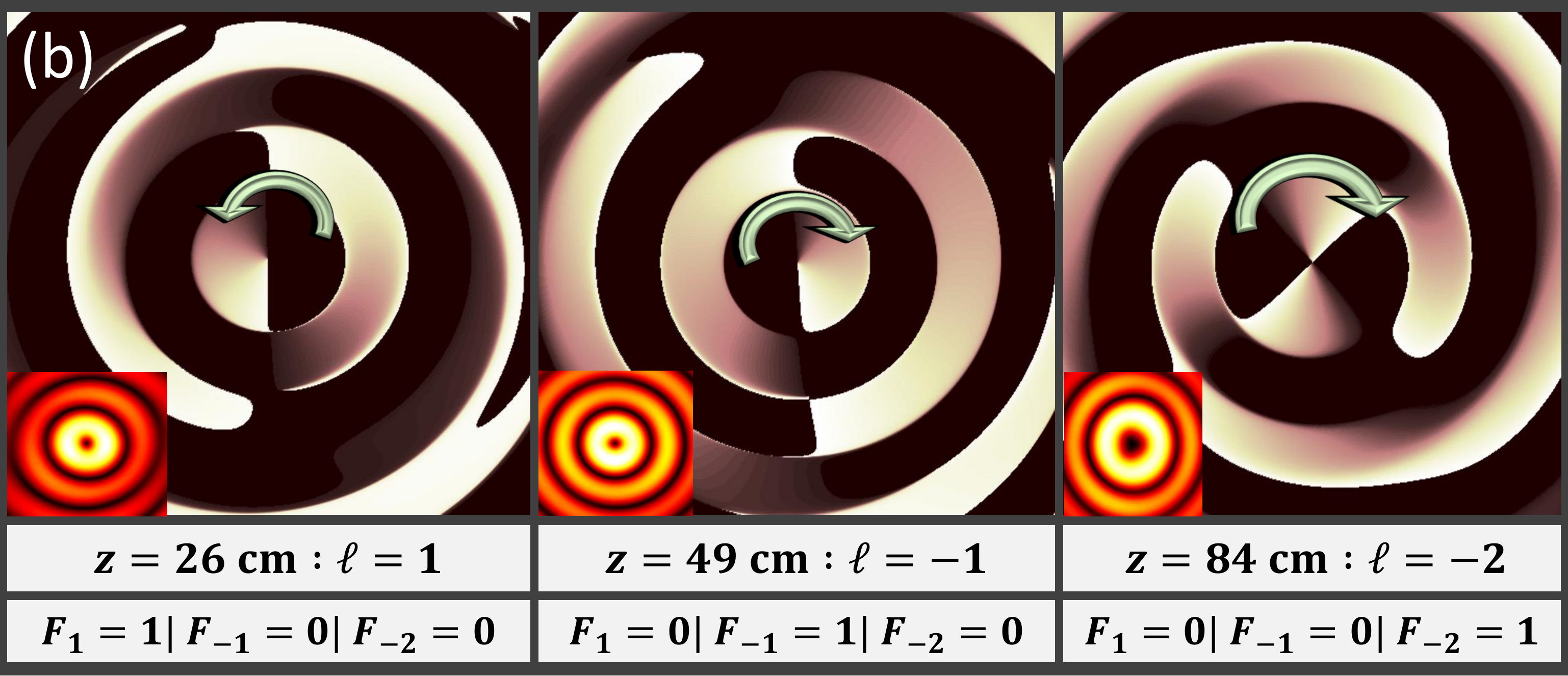}
	\caption{Phase-fronts that invert their handedness and increases their topological order with propagation. a) Schematic of helical phase whose topological charge takes the values $\ell= 1,-1,$ and $-2$ over three consecutive regions of space. The phase-front reverses its twist from $\ell=1$ to $\ell=-1$, and increases the number of its helices from $\ell=-1$ to $\ell=-2$. b) Evolution of the phase and intensity of a beam with 3 FW states ($\Psi = \psi_1 + \psi_{-1} + \psi_{-2}$). The phase reverses its sense of rotation (marked by the arrow) and increases its order of twist along the propagation direction. The insets depict the beam intensity profile where the beam radius increases upon the transition $\ell=-1$ to $\ell=-2$, implying an increase in the order of Bessel beams.}
	\label{Fig1}
\end{figure}

Longitudinal control of ($\ell$) can yield waveforms whose phase-fronts change their helicity with propagation. Figure~\ref{Fig1}(a) schematically illustrates a phase-front that reverses its helicity (transitioning from $\ell=1$ to $\ell=-1$) and doubles its number of helices (from $\ell=-1$ to $\ell=-2$) with propagation. This can be realized via superposition of FW states ($\Psi = \psi_{1} + \psi_{-1} + \psi_{-2}$) over length $L=1$ m, by incorporating the morphological function $F_{\ell}(z)$ 
\begin{equation}
\label{Eq20}
\begin{split}
F_{\ell}(z) \begin{cases}
F_1 = 1 \ \ \ \ \ \ \ \ 5 \ $cm$\leq z \leq 35 \ $cm$,    \\
F_{-1} = 1 \ \ \ \ \ \ 35 \ $cm$\leq z \leq 65 \ $cm$,    \\
F_{-2} = 1\ \ \ \ \ \    65 \ $cm$\leq z \leq 95 \ $cm$,    \\
F_{\ell} = 0|_{\forall \ell} \ \ \ \  $elsewhere$.
\end{cases}  
\end{split}
\end{equation}
A numeric evaluation of the phase and intensity profiles of $\Psi$, at three different propagation distances, is depicted in Fig.~\ref{Fig1}(b).
 
In order to demonstrate control of ($\ell$) along the beam axis by directly imaging the beam intensity profile rather than its phase, we add two waveforms each carrying multiple FW states such that 
\begin{equation}
\label{Eq0}
\Psi_{Total}(\rho,\phi,z,t) = \Psi_1 + \Psi_2 = \sum_{\ell_1=-\infty}^{\infty} \psi_{1,\ell_{1}} + \sum_{\ell_2=-\infty}^{\infty} \psi_{2,\ell_{2}}. 
\end{equation}
The two waveforms, $\Psi_1$ and $\Psi_2$, have slightly shifted longitudinal wavenumbers ($\Delta \zeta = \zeta_{1,\ell_1 m}-\zeta_{2,\ell_2 m}$) and arbitrary topological charges ($\ell_1$ and $\ell_2$) carrying opposite signs over finite space intervals. Accordingly, in the space interval corresponding to the superposition of FW states $\psi_{1,\ell_{1}}$ and $\psi_{2,\ell_{2}}$, a rotating beam structure is generated with angular orientation $\Phi_{\ell_{1},\ell_{2}}$ that varies with $z$ according to \cite{Ref17}
\begin{equation}
\label{Eq9}
\Phi_{\ell_{1},\ell_{2}} \propto \frac{z(\Delta \zeta) }{|\ell_{1}|+|\ell_{2}|}. 
\end{equation}
 
\begin{figure}[h!]
	\centering
	\includegraphics[width=.55\textwidth]{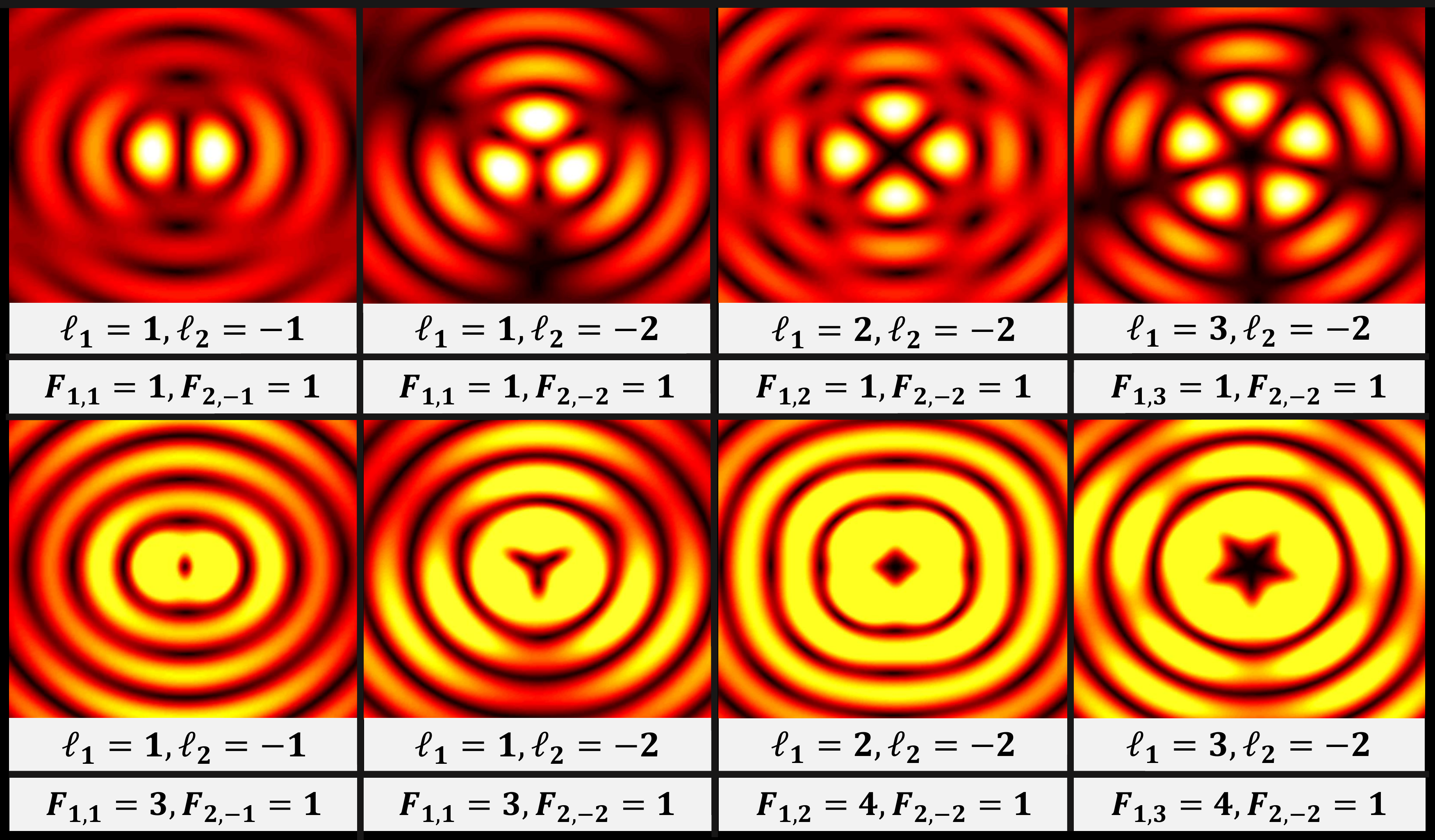}
	\caption{Transverse profiles generated by adding two waveforms with different topological charges ($\ell_{u}$) and by changing $F_{u,\ell_{u}}(z)$, where $u$ is the index of either of the two added waveforms in a given space interval ($u=1,2$). The top row depicts symmetric weights for the FW states; $F_{1,\ell_{1}}=F_{2,\ell_{2}}=1$, whereas the bottom row depicts asymmetric weights. The radial extent of the beam increases by increasing the value of $\ell_{u}$, implying an increase in the Bessel beams order.}
	\label{Fig2}
\end{figure}

The radial extent of the generated transverse profiles depends on ($\ell_{u}$) in $\psi_{u,\ell_{u}}|_{u=1,2}$ and the transverse beam shape is defined by $\ell_{u}$ and $F_{{u,\ell_{u}}}(z)|_{u=1,2}$. By controlling these parameters, the transverse intensity of the beam can be made to evolve into various profiles along the beam's propagation direction. Several topologies of rotating beam structures are illustrated in Fig.~\ref{Fig2} by tuning $\ell_{u}$ and $F_{{u,\ell_{u}}}(z)|_{u=1,2}$ as indicated on the figure.   

\begin{figure}[h!]
	\centering
	\includegraphics[width=0.8\textwidth]{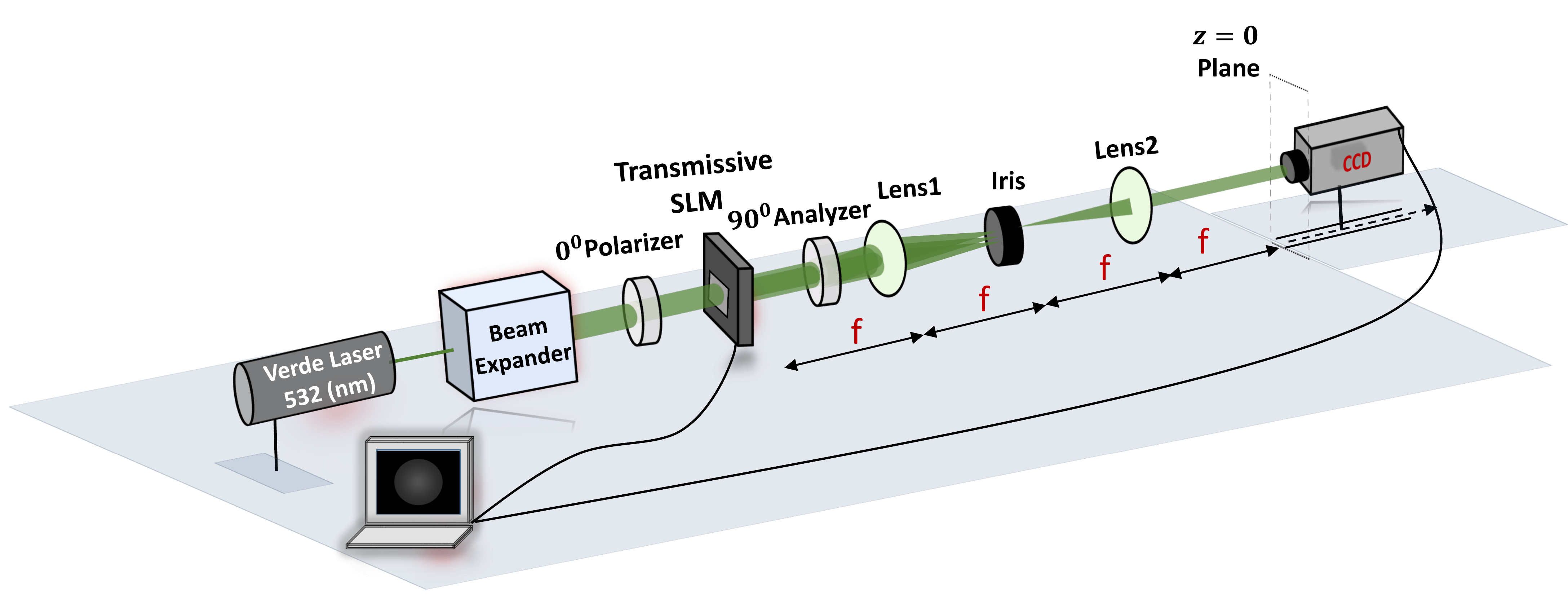}
	\caption{Experimental setup used to generate the twisted waveforms. A computer generated hologram is addressed onto a transmissive SLM that encodes the desired pattern on a 532 nm collimated laser beam. The SLM is sandwiched in a polarizer-analyzer configuration as it operates with maximum efficiency on vertically polarized incident light. The generated pattern is filtered and imaged using a 4-$f$ imaging system. The beam evolution is recorded using a CCD camera on a translation stage, where the $z=0$ plane lies in the focus of Lens2.}
	\label{Fig3}
\end{figure}

We experimentally demonstrate two different beam patterns by adding two waveforms each carrying multiple FW states according to Eq.~\ref{Eq0}. The experimental setup is depicted in Fig.~\ref{Fig3}. The waveforms in Eq.~\ref{Eq0} are calculated and transformed into a 2D computer generated hologram (CGH) that is mapped onto an amplitude Spatial Light Modulator (SLM) (please check the Methods section for more details on the CGH generation). The SLM encodes the hologram transmission function on 532 nm collimated laser beam. The resulting pattern is imaged and filtered using a 4-$f$ optical system and an iris that filters out the desired diffracted pattern from the on-axis noise \cite{Ref21,Ref22}. The evolution of the generated waveform is then recorded using a sliding CCD camera with 1 cm step resolution along the longitudinal direction.

We have experimentally generated two different patterns in which we set $N=7$. Accordingly, each FW state $\psi_{u,\ell_u}$ consists of $2N+1$= $15$ equal frequency Bessel beams of order $\ell_u$ with equally spaced longitudinal wavenumbers $Re\{\zeta_{\ell_{u},m}\} = Q_{u,\ell_u} + \frac{2\pi m}{L}$, where $Re\{\zeta_{\ell_{u},m}\}$ is the real part of  $\zeta_{\ell_u,m}$. The parameter $Q_{u,\ell_u}$ is a constant that defines the transverse beam localization \cite{Ref18} (also check Methods section). The longitudinal wavenumbers in $\Psi_1$ and $\Psi_2$ are centered around slightly shifted values $Q_{1,\ell_1}$ and $Q_{2,\ell_2}$; such that $\Delta \zeta = \Delta Q$ = 33 m$^{-1}$. 

\begin{figure*}[h!]
	\centering
	\includegraphics[width=.85\textwidth]{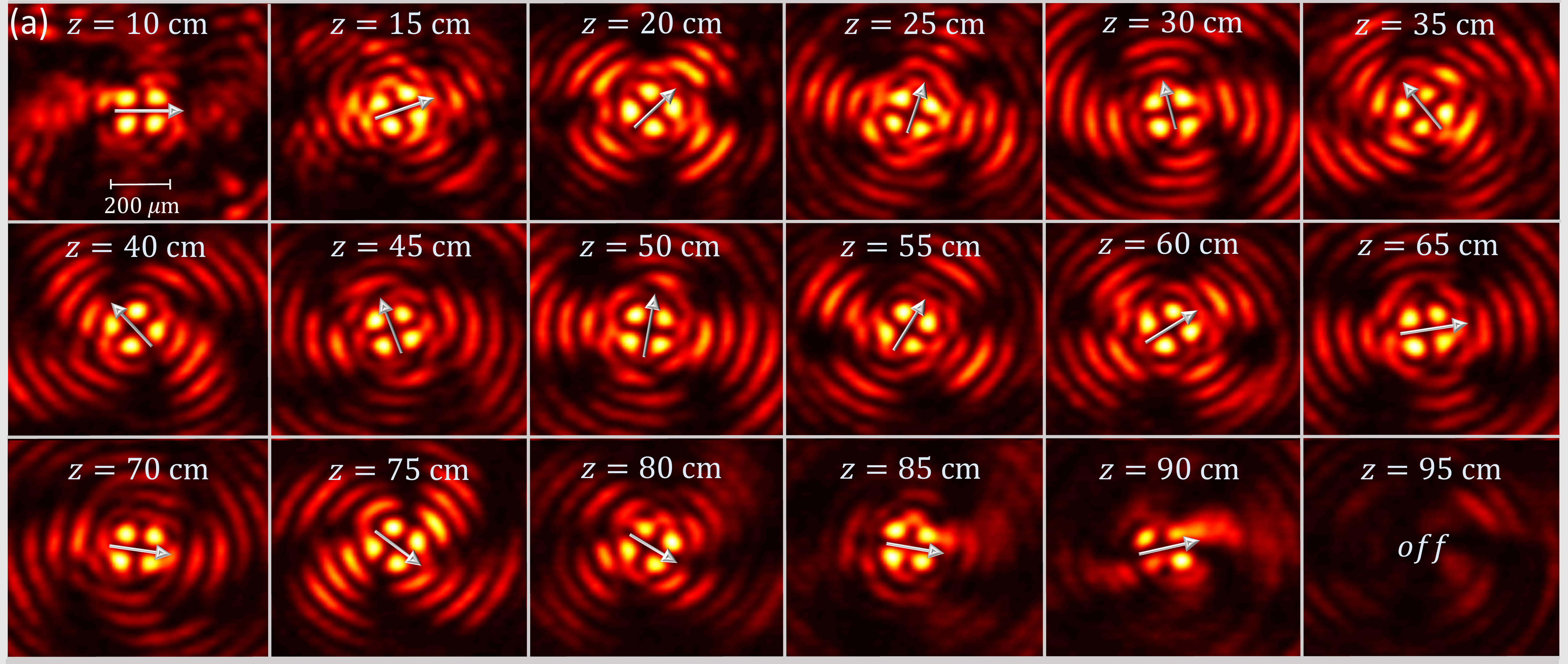}
	\includegraphics[width=.47\textwidth]{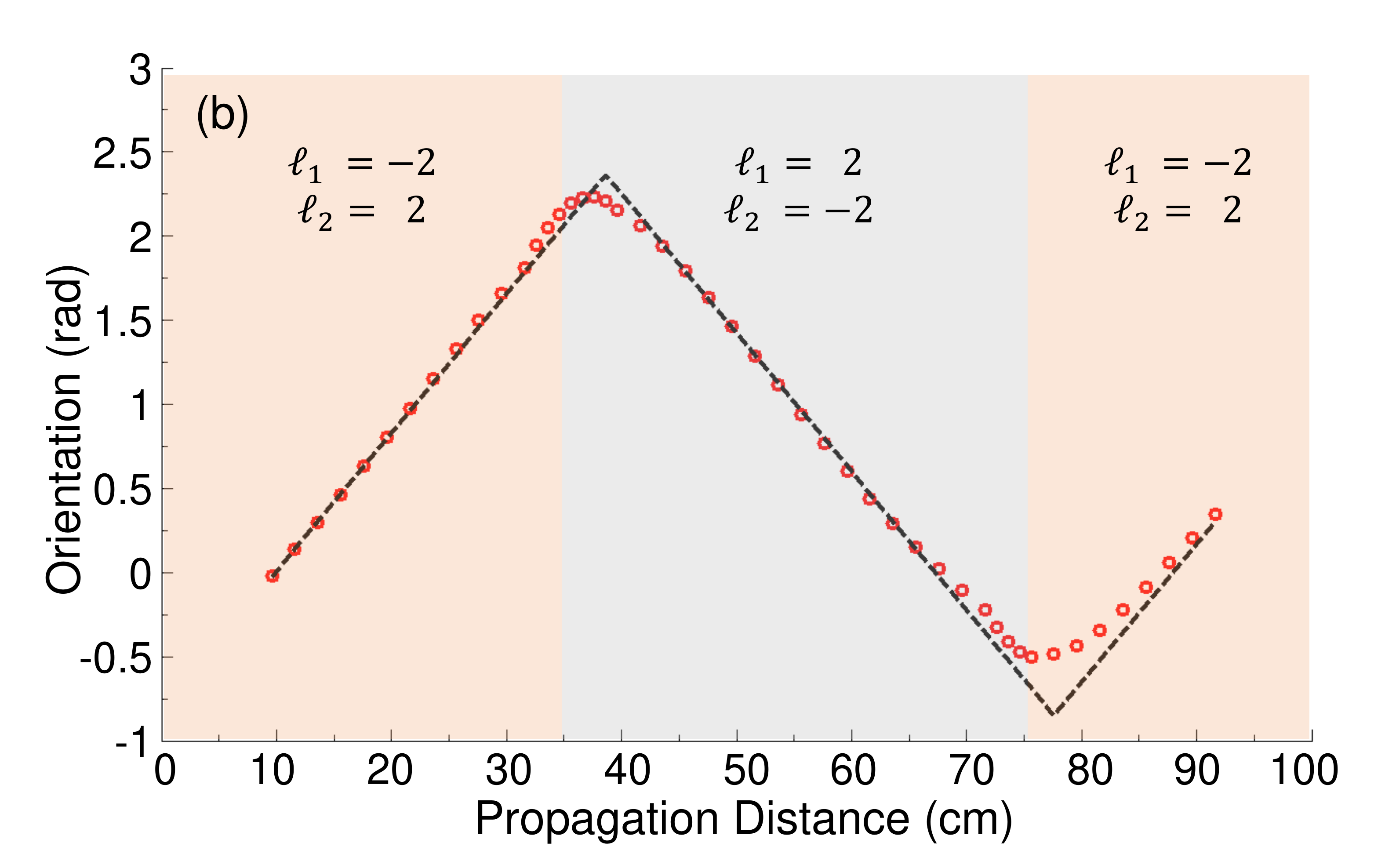}
	\includegraphics[width=.47\textwidth]{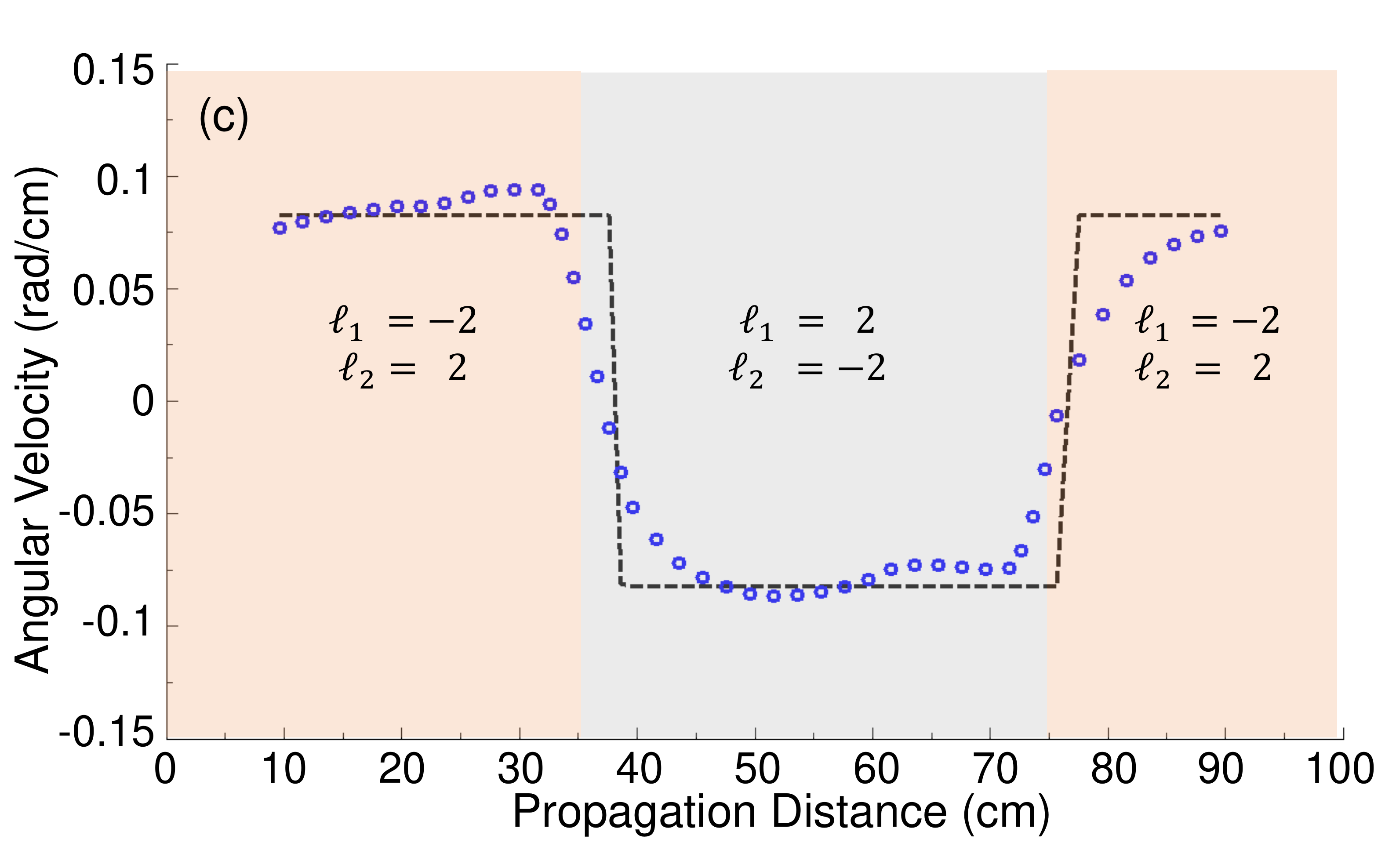}
	\caption{Evolution of the rotating intensity distribution $|\Psi_{Total}|^2$ with propagation. a) The four petals rotate in a counter clockwise direction over the range of $5$ cm $\leq z \leq 35$ cm. Inversion in the sense of rotation is observed in the middle row before the beam retains its initial sense of rotation within the range of  $75$ cm $\leq z \leq 90$ cm. The arrows are included to track the angular orientation of the beam. The full pattern dynamics can be viewed online in the videos: ``Supplementary Video 1'' and ``Supplementary Video 2''. b) Measured angular orientation of the beam (red circles) compared with theoretical prediction of Eq.\ref{Eq9} (black dashed lines). Clearly, the petals encounter linear rotation. The inversion of slope implies reversal of the sense of rotation. c) Angular velocity of the beam (blue circles) compared with theory (black dashed lines) calculated from the derivative of Eq.\ref{Eq9}. Negative velocities indicate inversion in the sense of rotation.}
	\label{Fig4}
\end{figure*}

In the first generated pattern, $\Psi_1$ and $\Psi_2$ consist of three FW states; $\ell_{1} = (-2,2,-2)$ and $\ell_{2} = (2,-2,2)$. The morphological function is given by
\begin{equation}
  \label{Eq7}
  \begin{split}
    F_{u,\ell_{u}}(z) \begin{cases}
    F_{1,-2} = F_{2,2} = 1 \ \ \ \ \ \ \  5 \ $cm$\leq z \leq 35 \ $cm$,    \\
    F_{1,2} = F_{2,-2} = 1 \ \ \ \ \ \ \ 35 \ $cm$\leq z \leq 75 \ $cm$,    \\
    F_{1,-2} = F_{2,2} = 1\ \ \ \ \ \ \          75 \ $cm$\leq z \leq 90 \ $cm$,    \\
  F_{u,\ell_{u}} = 0|_{\forall u,\ell_u} \ \ \ \ \ \ \ \ \ \ \ \ \      $elsewhere$.
  \end{cases}  
  \end{split}
  \end{equation} 

According to the above $F_{u,\ell_{u}}(z)$, the generated beam should posses four revolving petals that invert their sense of rotation twice: at $z=35$ cm and again at $z=75$ cm. The evolution of the transverse profile of the resulting rotating beam is recorded via a CCD camera over a range of 1 m and is shown in Fig.~\ref{Fig4}(a). The small arrows are superimposed on the recorded images to track the sense of rotation of the beam. As suggested by Eq.~\ref{Eq7}, the beam exhibits a counter clockwise rotation before it inverts its sense of rotation as depicted in the middle row of Fig.~\ref{Fig4}(a). Afterwards, at the distance of 75 cm, the beam recovers its initial rotation direction. The angular orientation of the rotating petals is plotted in Fig.~\ref{Fig4}(b). Clearly, the curve inverts its slope twice over the range of 35 cm $\leq z \leq$ 75 and 75 cm  $\leq z \leq$ 90 cm, signifying a reverse in the sign of the topological charge. Moreover, the rotating petals exhibit an angular velocity $\frac{\partial \Phi}{\partial z} \sim$ 0.0825 rad/cm, in agreement with Eq.~\ref{Eq9}. This angular velocity is also depicted in Fig.~\ref{Fig4}(c) showing acceleration and deceleration associated with changing the rotation direction, as expected. 

In the previous scenario, we demonstrated the possibility of inverting the sign of the effective topological charge as the beam propagates. In the following we show the possibility of controlling the sign and absolute value of the effective topological charge along the beam axis. In this scenario, $\Psi_1$ and $\Psi_2$ have four FW states such that $\ell_{1} = (-1,-2,0,3)$ and $\ell_{2} = (2,3,0,-4)$. 
\begin{figure*}[h!]
	\centering
	\includegraphics[width=.85\textwidth]{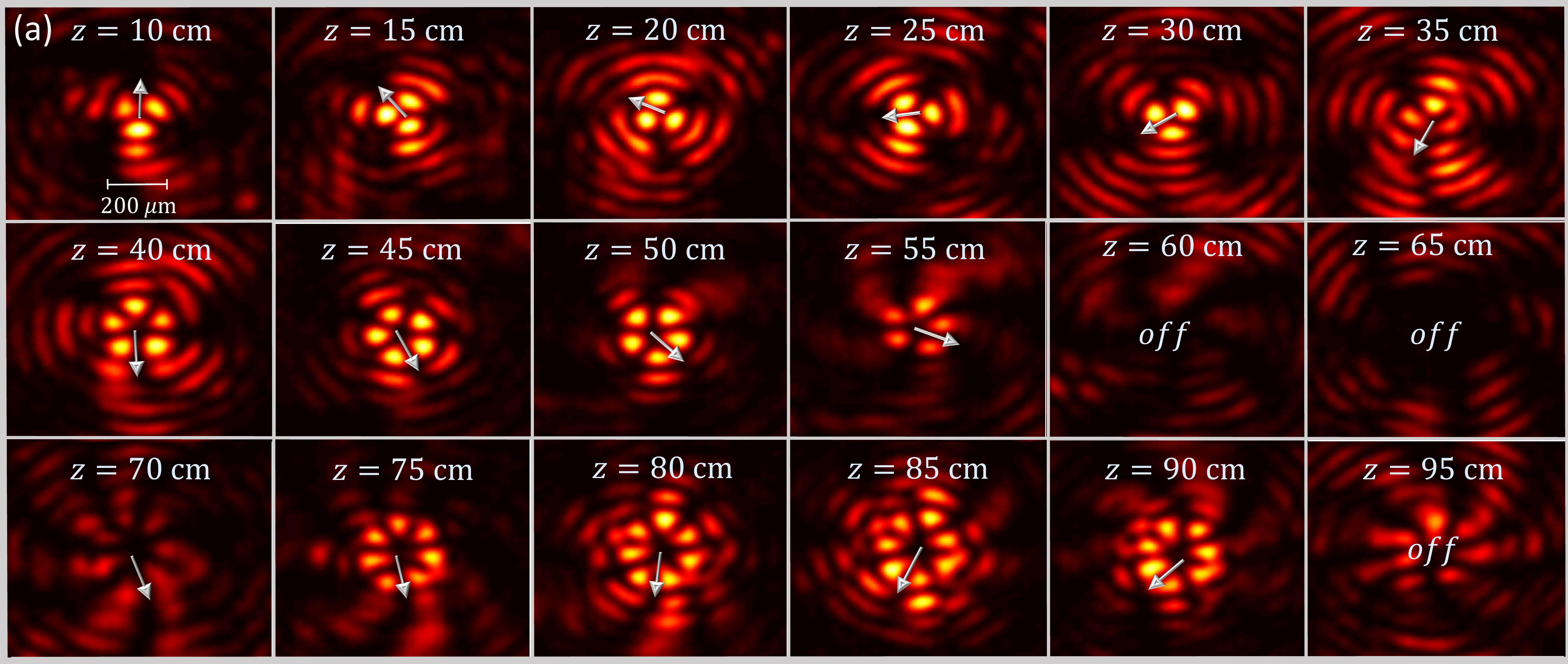}
	\includegraphics[width=.47\textwidth]{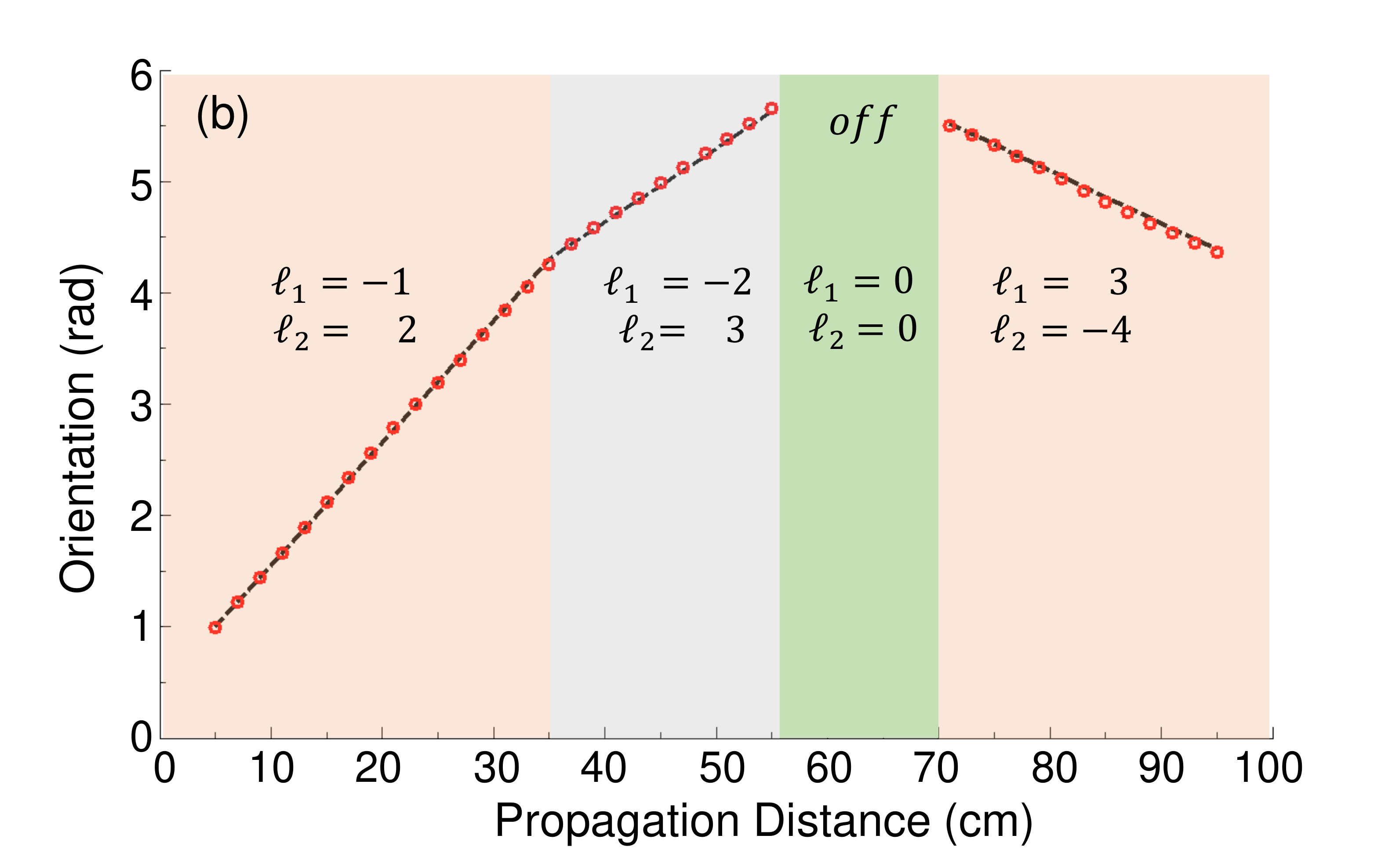}
	\includegraphics[width=.47\textwidth]{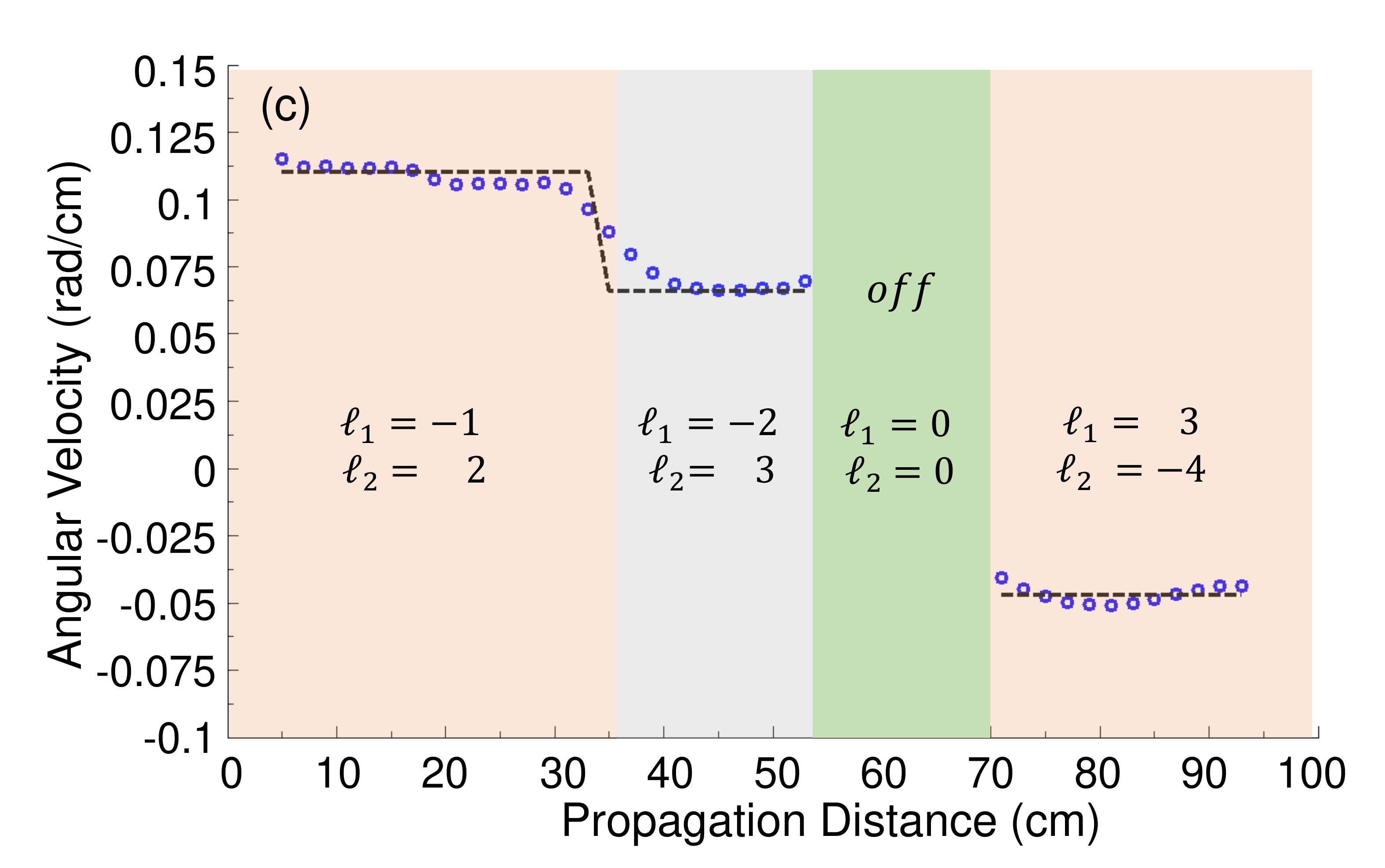}
	\caption{Intensity pattern of the twisted light generated from FWs. a) Three petals rotate with a counter clockwise sense of rotation over the range of 5 cm $\leq z \leq$ 35 cm before it evolves to five petals, indicating an increase in its phase twist. Afterwards, the petals carry zero intensity between $55$ cm $\leq z \leq 70$ cm. The beam then evolves in the form of seven petals rotating in the clockwise direction. The full pattern dynamics can be viewed online in the videos: ``Supplementary Video 3'' and ``Supplementary Video 4''.  b) Measured angular orientation of the rotating beam (red circles) in comparison with theory (black dashed line) calculated from Eq.~\ref{Eq9}. The slope signifies the rate of rotation (angular velocity) and its sign indicates the sense of rotation. c) Measured (blue circles) and simulated (black dashed line) angular velocity of the beam, also calculated from Eq.~\ref{Eq9}, with propagation distance. It is observed that the angular velocity is reduced when the number of the rotating petals increased.}
	\label{Fig5}
\end{figure*}

The morphological function is chosen as  
\begin{equation}
\label{Eq8}
  \begin{split}
  F_{u,\ell_{u}}(z) \begin{cases}
  F_{1,-1} = F_{2,2}  = 1 \ \ \ \ \ \ \  5 \ $cm$\leq z \leq 35 \ $cm$,    \\
  F_{1,-2} = F_{2,3}  = 1 \ \ \ \ \ \ \  35 \ $cm$\leq z \leq 55 \ $cm$,    \\
  F_{1,0}  = F_{2,0}  = 0 \ \ \ \ \ \ \ \ \ 55 \ $cm$\leq z \leq 70 \ $cm$,    \\
  F_{1,3}  = F_{2,-4} = 1 \ \ \ \ \ \ \  70 \ $cm$\leq z \leq 95 \ $cm$,    \\
  F_{u,\ell_{u}} = 0|_{\forall u,\ell_{u}} \ \ \ \ \ \ \ \ \ \ \ \ \      $elsewhere$.
  \end{cases}  
  \end{split}
\end{equation} 

According to Eq.~\ref{Eq8}, the generated beam carries three petals that rotate with a linear speed in the counterclockwise direction over the range of 5 cm $\leq z \leq$ 35 cm. The beam then evolves into five petals rotating in the same direction for 35 cm $\leq z \leq$ 55 before petals carry zero intensity in the range of 55 cm $\leq z \leq $ 70 cm. The beam then evolves into a rotating structure comprised of seven petals with an opposite sense of rotation (clockwise direction), signifying both an increase and inversion in its azimuthal phase twist. 

The evolution of intensity distribution $|\Psi_{Total}|^2$ is depicted in Fig.~\ref{Fig5}(a). The angular orientation and velocity of the rotating petals are plotted in Figs.~\ref{Fig5} (b) and (c), respectively; showing very good agreement with theoretical predictions. The rotating beam structure initially exhibits larger angular velocity (when it carries three petals) as compared to the case when it evolves into five and seven petals, in full agreement with Eq.~\ref{Eq9}. Clearly, the radial extent of the beam increases when its effective topological charge increases, as dictated by Helmholtz wave equation. The optical beams with dynamic control over their OAM demonstrated in Figs.~\ref{Fig4} and \ref{Fig5} are a result of careful superposition of several FW states. In essence, in a given space interval, only the FW states \textit{selected} by the morphological function $F_{u,\ell_{u}}(z)$ effectively contribute to the beam while the contributions of all other FW states vanish by being stored away from the center of the beam. 

Finally, few remarks are in order: first, although we have demonstrated dynamic control over OAM modes with constant angular velocity, beams with accelerated rotation can be realized by incorporating nonlinear azimuthal phase in the FW states ($\psi_{\ell}$) \cite{Ref17}. Second, while the experimental beams presented herein are generated in lossless medium (air); nevertheless, attenuation-resistant beams with OAM can also be generated in absorbing media by engineering $F(z)$ such that it carries an inverse of the medium loss profile, as discussed in \cite{Ref19c,Ref22b}. Finally, it is worth noting that FWs possess additional compelling features such as self-reconstruction and diffraction resistance which are inherent characteristics of Bessel beams \cite{Ref23}.

In conclusion, we have demonstrated advanced control over OAM modes by changing the sign and value of the topological charge ($\ell$) as the beam propagates. These degrees of flexibility in manipulating the beam can be utilized in applications such as optical trapping, dense data communications, and imaging, to name a few. We expect that these utilizations and advances will open new directions in the field of optical science and its applications. 
\section*{Methods}

\subsection*{Theoretical Model}

In the theoretical formulation of FWs in general absorbing media with complex index of refraction ($n=n_r + in_i$), $k$ is a complex wavenumber described by $k = k_r + i k_i$. Here, considering a single FW state $\psi_{\ell}$, the real part of the longitudinal wavenumber of the $m^{th}$ Bessel beam is $Re\{\zeta_{\ell m}\} = Q_{\ell} + \frac{2\pi m}{L}$. The parameter $Q_{\ell}$ is a constant that defines the transverse localization of the field. In addition, we have required the transverse wavenumbers to acquire real values in order to exemplify the case of generating a beam in lossless material before it enters a medium with arbitrary index of refraction $n$. Accordingly, the imaginary part of the longitudinal wavenumber is expressed as $Im\{\zeta_{\ell m}\} = \frac{\omega^2}{c^2}\frac{n_r n_i}{Re\{\zeta_{\ell m}\}}$ and the transverse wavenumber is given by
\begin{equation}
\label{Eq2}
\eta_{\ell m} = \sqrt{(n_r^2-n_i^2)\frac{\omega^2}{c^2} - (Q_{\ell} + \frac{2\pi m}{L})^2 + \left(\frac{\omega^2}{c^2}\frac{n_r n_i}{Q_{\ell} + \frac{2\pi m}{L}}\right)^2}.
\end{equation}

To enforce positive values for the real part of the longitudinal wavenumber and real values for the transverse wavenumebrs ($\eta_{\ell m}$), the following condition must be satisfied
\begin{equation}
\label{Eq3}
0\leq Re\{\zeta_{\ell m}\} = Q_{\ell} + \frac{2\pi m}{L} \leq \frac{\omega}{c} \sqrt{\frac{(n_r^2-n_i^2)+\sqrt{(n_r^2-n_i^2)^2 + 4n_r^2 n_i^2}}{2}}.
\end{equation}

\subsection*{Numerical Parameters}

For our numerical simulations and computer generated holograms (CGHs) we have adopted the following numerical values: We have assigned a value of $0.9999958 \times k$ to the $Q_{1,\ell_1}$ parameter (for all the FW states in $\Psi_1$). By substituting in Eq.~\ref{Eq3}, this results in a maximum allowable value of $N$ equal to $7$, resulting in $2N+1 = 15$ Bessel beams in the superposition. The $Q_{2,\ell_2}$ parameter, associated with FW states of $\Psi_2$, is set to $0.999993 \times k$. This yields a maximum value of $N=13$. Accordingly, we have selected $N=7$ for both waveforms. Given the laser source wavelength $\lambda=532$ nm, in that case $\Delta \zeta = \Delta Q = 33$ m$^{-1}$. These values of $Q_{u,\ell_u}$ are chosen such that the generated Bessel beams possess transverse wavenumbers ($\eta_{\ell m}$) that are compatible with the spatial bandwidth of our SLM. The radial extent of each generated FW state ($\rho_{\ell}$) can be estimated by solving $\frac{\partial}{\partial \rho} J_{\ell}(\rho \sqrt{\omega^2/c^2 - Q_{\ell}^2})|_{\rho=\rho_{\ell}}=0$. 

\subsection*{Computer Generated Hologram}
In our experiment, we have used an amplitude mask to express the complex transmission function of the waveform at the $z=0$ plane; i.e. to define $\Psi_{Total}(\rho,\phi,z=0,t)$. The hologram equation can be mathematically expressed as
\begin{equation}
\label{Eq5}
H(x,y) = \frac{1}{2}\lbrace\beta(x,y)+\alpha(x,y)cos[\phi(x,y)-2\pi(u_0x + v_0y)]\rbrace,
\end{equation} 
where, $\alpha(x,y)$ and $\phi(x,y)$ are the amplitude and phase of $\Psi_{Total}(\rho,\phi,z=0,t)$, respectively. A bias function $\beta(x,y)$ is chosen as a soft envelope for the amplitude $\alpha(x,y)$ and is given by $\beta(x,y) = [1+\alpha(x,y)^2]/2$ \cite{Ref24}. In addition, the pattern is interfered with a plane wave $exp[2\pi i(u_0x + v_0y)]$. This shifts the encoded pattern off-axis to the spatial frequencies ($u_0,v_0$) in the Fourier plane; thus making it easier to filter out the shifted pattern from the undesired on-axis noise by using an iris. The parameters $u_0$ and $v_0$ were set to $1/(4\Delta{x})$; where $\Delta{x}$ is the SLM pixel pitch (36 $\mu$m). The shifted beam is then imaged using a 4-$f$ optical system, with focal lengths of 20 cm. 

In principle, our theoretical model yields periodic waveforms (with infinite power flux) \cite{Ref20}. As such, a finite truncated circular aperture is superimposed over the generated hologram. For the efficient generation of the desired waveform over a spatial range $L$, a sufficient condition for the aperture diameter ($D$) is
\begin{equation}
\label{Eq6}
D \geq 2L \sqrt{(\frac{k}{Re\{\zeta_{\ell, m=-N}\}})^2-1}.
\end{equation} 


\end{document}